\documentclass[aps, pra, 10pt, a4paper, reprint, nofootinbib, superscriptaddress]{revtex4-2}
\usepackage[utf8]{inputenc}
\usepackage[sc,osf]{mathpazo}
\usepackage[T1]{fontenc}
\usepackage{amsfonts}
\usepackage{amssymb}
\usepackage{amsmath}
\usepackage{amsthm}
\usepackage{graphics}
\usepackage{graphicx}
\usepackage{braket}
\usepackage[colorlinks=true,linkcolor=blue,urlcolor=blue,citecolor=blue]{hyperref}

\newtheorem{definition}{Definition}

\usepackage{color,soul}
\usepackage{xcolor}

\begin{document}
\title{Quantum entanglement percolation under a realistic restriction}

\author{Shashaank Khanna}
\affiliation{Discipline of Physics, Indian Institute of Technology Indore, Khandwa Road, Simrol, Indore 453 552, India}
\affiliation{Harish-Chandra Research Institute, HBNI, Chhatnag Road, Jhunsi, Allahabad 211 019, India}

\author{Saronath Halder}
\affiliation{Harish-Chandra Research Institute, HBNI, Chhatnag Road, Jhunsi, Allahabad 211 019, India}

\author{Ujjwal Sen}
\affiliation{Harish-Chandra Research Institute, HBNI, Chhatnag Road, Jhunsi, Allahabad 211 019, India}

\begin{abstract}
The problem of establishing Bell and Greenberger-Horne-Zeilinger states between faraway places or distant nodes of a circuit is a difficult and an extremely important one, and a strategy which addresses it is entanglement percolation. We provide a method for attaining the end through a quantum measurement strategy involving three-, two-, and single-qubit measurements on a single-layer honeycomb lattice of partially entangled pure bipartite entangled states. We then move over to a double-layered lattice, and  introduce entanglement percolation on that lattice under a realistic restriction on local quantum operations and classical communication allowed on the nodes of the lattice. When applied to a single-layered honeycomb lattice, our strategy would call for less noise effects in an actual realization than when the same phenomenon is attained via existing methods. Moreover, for the double-layered honeycomb lattice, we report advantage of quantum entanglement percolation over classical entanglement percolation under the realistic restriction.  
\end{abstract}
\maketitle

\section{Introduction}\label{sec1}
One of the major problems in quantum information \cite{Nielsen00} is of distributing entangled states \cite{Horodecki09, ent-2, ent-3}. Entanglement is often a fragile resource and decoherence tends to frequently make the problem of distributing entanglement a difficult one. But distributing entanglement, be it between two remote positions on a lattice or between two stations separated by a relatively large distance, can have a multitude of uses, ranging from  quantum computers \cite{qcompu} to quantum key distribution \cite{BB84, Ekert91, HMP, Elliott02} and quantum dense coding \cite{Ben92, DDC, DML, PAAU, PAU, TRAU, STTA}. Quantum networks \cite{netw1, netw2, netw3, netw4, netw5, netw6, netw7, netw8} have been employed as a solution to the problem of distributing entanglement. They consist of nodes where each node can station many qubits. Different qubits at a particular node can be entangled with other qubits at other nodes. A network often has a well-defined geometric structure, and may  form a lattice, for example, a triangular or a square lattice. Local quantum operations at the nodes and classical communication between the nodes are usually accessible in a realistic situation, and are thereby adopted in theoretical considerations of manipulation of the structure and  connectivity of a quantum network.

An associated problem is that of establishing maximally entangled states (also called Bell states) between two faraway places. Using maximally entangled states for different protocols like quantum teleportation \cite{Ben93} and quantum cryptography \cite{Gisin02} is extremely important, since they often provide the maximum advantage over the corresponding classical protocols. (See \cite{not-maximal, ADGL, HAU, ASAU, RDDSS, SASAU, CSAU, not-maximal-coherent} however.) 

Nielsen's majorization criterion answered the question whether a single copy of a bipartite pure state can be converted to another, deterministically and under local operations and classical communication (LOCC) \cite{Niel99}. Vidal derived the complete set of monotones for local pure state transformations and found the formula for the maximum probability of successfully converting -- under LOCC -- a bipartite pure state to another \cite{Vidal99} (see also \cite{Lo-Popescu, Hardy, Jonathan-Plenio}). The entanglement swapping scheme was earlier introduced by {\.Z}ukowski \emph{et al.} in Refs.~\cite{Zuko93} (see also \cite{MAZH, output}). Bose, Vedral, and Knight generalized this scheme to a multi-particle scenario and applied it to a communication network \cite{Bose98}. 

Ac{\' i}n, Cirac, and Lewenstein (ACL) \cite{Acin07} developed the ``classical entanglement percolation'' (CEP) protocol and used it to show how the concept of percolation \cite{Grimmett99} in statistical mechanics can be applied in the context of sharing quantum entanglement. (See \cite{Perseguers08, Lap09, Perseguers10, Perseguers13, Daknam-e-Deke-brisTir-dupure} for further studies.) Precisely, they showed how it can be utilized to achieve the task of distributing entanglement between  faraway nodes on a network and to establish a maximally entangled Bell state between two distant nodes of an asymptotically large lattice. They further developed a protocol which they termed as ``quantum entanglement percolation'' (QEP), and showed how QEP can succeed where CEP could not, in accomplishing the task of entanglement distribution. 

In this paper, we will mainly be concerned with QEP, and use shared pure entanglement for this purpose. In particular, a quantum measurement strategy is constructed which helps to establish maximally entangled states between two ``antipodal'' end-nodes of a lattice. We discuss about both single- and double-layered honeycomb lattices where the present measurement strategy can be utilized. The hexagonal (or honeycomb) lattice is one of the most widely-used lattices in the theory and experiment of ultra-cold gases and condensed matter physics (see e.g.~\cite{madhup1, madhup2, madhup3, madhup4, madhup5} for some recent examples). We believe that this proliferation of hexagonal lattices in real systems would lend the strategy reported in the paper to be more prone to being realized when considered for that lattice. Apart from this practical aspect, there is a mathematical tractability issue that led to the choice of the lattice. The percolation threshold can be analytically calculated for the honeycomb lattice. As a result, the results obtained in the manuscript are independent of numerical approximations, in the context of calculation of percolation thresholds. Only for a few lattices can the percolation threshold probability be calculated exactly, with honeycomb and triangular lattices being two such lattices. This has led us, and may have led the former works in the literature, to choose to study transformations from a honeycomb to a triangular lattice. 

Our strategy involves three-, two-, and single-qubit measurements. The outcome is in a Greenberger-Horne-Zeilinger state~\cite{eiTa-GHZ, NDMA} between an arbitrary number of nodes of the lattice, which can then be transformed to a maximally entangled two-qubit states between two faraway nodes. Sticking to a single layer of the honeycomb lattice, we compare our strategy with previous entanglement percolation strategies in the literature with respect to the resources utilized, and show that the number of measurements required for the success of our strategy is less than those required in both Ref.~\cite{Acin07} and Ref.~\cite{Perseguers10}. It is thus plausible that our measurement strategy would have less noise effects in an actual experimental realization compared to the same based on the strategies in Refs.~\cite{Acin07,Perseguers10}. When applied to a single-layered honeycomb lattice, it may look like our measurement strategy is not that effective since there is no advantage of QEP over CEP. But to find out the non-trivial feature(s) of our measurement strategy, we apply it to a double-layered honeycomb lattice with the restriction that the layers are provided one by one, i.e., no joint measurement on both layers is allowed. In this way, by introducing entanglement percolation under what we refer to as ``restricted'' local quantum  operations and classical communication (rLOCC), we show how QEP can be advantageous over CEP via our measurement strategy. We note that the restriction we are talking about can be experiment friendly. Because compared to full LOCC, rLOCC will be easier to implement. In fact, the main motivation of considering rLOCC comes from the adaptive strategies which have drawn significant attention recently; for example, see \cite{Harrow10, Higgins11, Katariya21, Banik21, Pauwels22}.  

The rest of the paper is arranged in the following way. In Sec.~\ref{sec2}, we provide a recapitulation of a few tools that will be necessary for our analysis. Thereafter, in Sec.~\ref{sec3}, we present the  main results, and finally, in Sec.~\ref{sec4}, we provide the concluding remarks.

\section{Collecting the tools}\label{sec2}
\subsection{Classical entanglement percolation}
\begin{figure}[t!]
\includegraphics[scale=0.08]{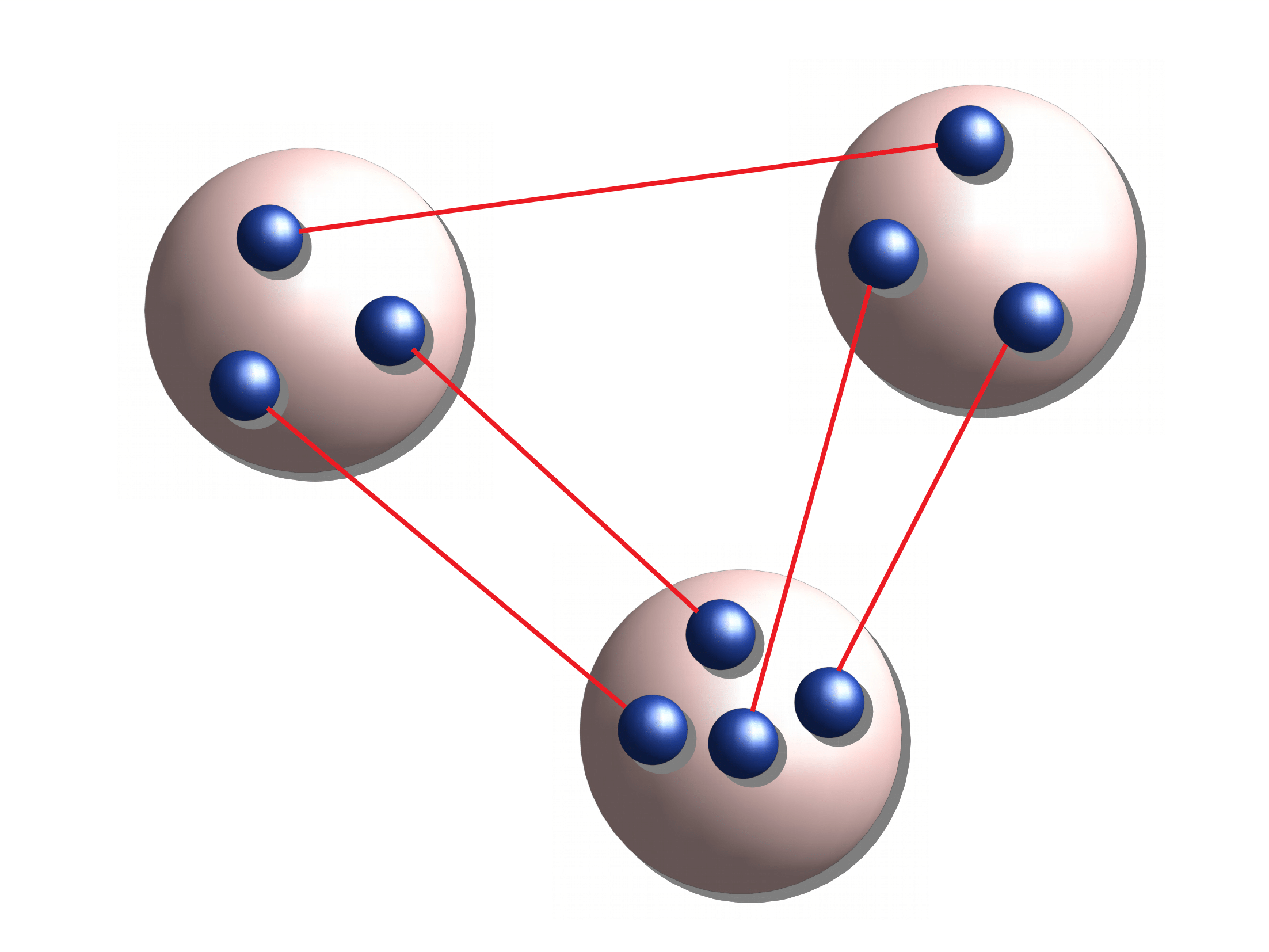}
\caption{Schematic diagram of a quantum network. The network is formed by a collection of nodes, each of which contains a cluster of qubits. A qubit in one node is typically entangled with a qubit in a different node. The smaller circles represent the qubits, while the larger ones represent the nodes. The lines represent the entangled states.}\label{fig1}
\end{figure}

We begin with a description of the protocol for CEP. Firstly, any node of a lattice can contain any number of qubits and secondly, the qubits of two different nodes can be connected via partially entangled states. See Fig.~\ref{fig1}, where the nodes within a quantum network are shown. The geometry created due to these partially entangled qubits at different nodes, forms the structure of the lattice. The protocol begins by applying LOCC between the nodes to convert the partially entangled states to maximally entangled states. After this, some of the previous links are broken and the probability that an initial partially entangled state is converted to a maximally entangled one, is governed by the singlet conversion probability (SCP) of the initial states.

\begin{figure}[t!]
\includegraphics[scale=0.75]{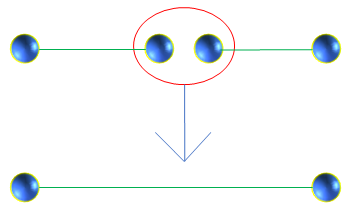}
\caption{Entanglement swapping. Entanglement swapping, probably the earliest and the simplest quantum network, can be seen as  a method of entangling two quantum systems that have never met, but are each entangled with two further quantum systems who have interacted in the past. In the schematic given, the quantum systems are represented by the blue circles and the lines represent entangled states. The red ellipse enclosing the two blue circles indicate an interaction, possibly via a measurement, between the two blue circles. In the figure, the two blue circles at the extremes get entangled by the interaction. The arrow represents the flow of operations in time.}\label{fig2}
\end{figure}

Now for every lattice, there exists a percolation threshold which is the critical value of the occupation probability in the lattice, such that infinite connectivity (percolation) occurs. In CEP, if the SCP is greater than the percolation threshold for the given lattice, then an infinite cluster forms in the lattice. This infinite cluster consists of nodes which are all linked with maximally entangled states and thus, one finds many paths along which one can do entanglement swapping (see Fig.~\ref{fig2}) to create a maximally entangled state between two faraway nodes of the given lattice. This can be performed, provided the two such nodes lie in the same cluster, the probability of which is $\theta(p)$, which is strictly greater than zero if SCP is greater than the percolation threshold of the lattice. So, the task of creating a maximally entangled state between two end nodes of the lattice has been accomplished with a strictly non-vanishing probability. The same would not have been possible using only entanglement swapping (without the singlet conversion step in CEP), since in that case, as was shown in Ref.~\cite{Acin07}, if the initial states were partially entangled, then the probability of succeeding would have decayed exponentially, with increasing lattice distance between the nodes.

\subsection{Quantum entanglement percolation}
In QEP, the original lattice structure, using some particular quantum measurements, is converted to some other lattice for which the percolation threshold is lower than that of the parent lattice. Thenceforth, CEP  is applied on the new lattice. To demonstrate the effectiveness of their protocol, ACL  used a double-layered honeycomb lattice in which percolation is not possible as the critical amount of entanglement (which is governed by the SCP here) is less than the percolation threshold \cite{Acin07}. Carrying out measurements in the Bell basis at the nodes, they converted their original lattice structure to a triangular lattice which has a lower percolation threshold, and thus meets the criterion, of the critical amount of entanglement being greater than the percolation threshold, for entanglement percolation to succeed in the new lattice. The Bell basis is given by the set of four orthonormal states, \((1/\sqrt{2})(|00\rangle \pm |11\rangle)\), \((1/\sqrt{2})(|01\rangle \pm |10\rangle)\). As evident, this protocol of entanglement percolation uses the richness of the geometry of two-dimensional lattices. Further, the particular lattice transformation used is one of the most important factors leading to the success of the QEP. It is important to note here that a double-layered honeycomb lattice was used, since using the measurement strategy described in Ref.~\cite{Acin07}, it is not possible to convert the single-layered honeycomb lattice to a triangular lattice.

In this paper, we will present a quantum measurement strategy which helps to establish multiparticle genuine entangled states between an arbitrarily large number of nodes and maximally entangled states between two end nodes of the lattice by using a single-layered honeycomb lattice. We then apply our measurement strategy to a double-layered honeycomb lattice.

\subsection{Schmidt decomposition}
If $|\psi\rangle$ is a pure state which belongs to a bipartite quantum system, described on the Hilbert space $\mathcal{H}$ = $\mathcal{H}_A\otimes\mathcal{H}_B$, then there exist orthonormal bases $\{|i_A\rangle\}$ and $\{|i_B\rangle\}$ in $\mathcal{H}_A$ and $\mathcal{H}_B$ respectively, such that
\begin{equation}\label{eq1}
|\psi\rangle = \sum\limits_{i}\sqrt{\alpha_i}|i_A\rangle|i_B\rangle,
\end{equation}
referred to as the Schmidt decomposition of \(|\psi\rangle\), where $\sqrt{\alpha_i}$ are non-negative real numbers which satisfy the condition $\sum_i\alpha_i = 1$. The $\sqrt{\alpha_i}$ are known as Schmidt coefficients.

\subsection{Nielsen's majorization criterion}
Nielsen found the necessary and sufficient condition that a pure entangled state $|\psi\rangle$ can be deterministically converted into another pure entangled state $|\phi\rangle$ under LOCC \cite{Niel99}. Consider a pure entangled state $|\psi\rangle\in\mathbb{C}^n\otimes\mathbb{C}^n$. Let the Schmidt decomposition of the state $|\psi\rangle$ be given by
\begin{equation}\label{eq2}
|\psi\rangle = \sum\limits_{i=1}^{n}\sqrt{\alpha_i}|i_A\rangle|i_B\rangle,
\end{equation}
where $\sum_{i=1}^{n}\alpha_i$ = $1$ and $\alpha_i\geq\alpha_{i+1}\geq0$. The problem is to find whether it can be converted, exactly and deterministically at the level of a single copy and under LOCC, to another pure state $|\phi\rangle\in\mathbb{C}^n\otimes\mathbb{C}^n$, whose Schmidt decomposition is given by
\begin{equation}\label{eq3}
|\phi\rangle = \sum\limits_{i=1}^{n}\sqrt{\beta_i}|i_A\rangle|i_B\rangle,
\end{equation}
where $\sum_{i=1}^{n}\beta_i = 1$ and $\beta_i\geq\beta_{i+1}\geq0$. Let us  define the Schmidt vectors, $\lambda_{\psi} = (\alpha_1, \alpha_2, \ldots, \alpha_n)$ and $\lambda_{\phi} = (\beta_1, \beta_2, \ldots, \beta_n)$. Then the Nielsen's criterion tells us that $|\psi\rangle$ can be converted to $|\phi\rangle$ under LOCC if and only if $\lambda_{\psi}$ is majorized by $\lambda_{\phi}$ (written as $\lambda_{\psi}$ $\prec$ $\lambda_{\phi}$), that is, iff
\begin{equation}\label{eq4}
\sum\limits_{i=1}^k\alpha_i\leq\sum\limits_{i=1}^k\beta_i
\end{equation}
for all $k = 1, 2, \ldots, n$.

\subsection{Singlet conversion probability}
Vidal \cite{Vidal99} showed that the pure state  $|\psi\rangle\in\mathbb{C}^n \otimes\mathbb{C}^n$ can be locally converted to the pure state \(|\phi\rangle\) of the same Hilbert space with a maximum probability given by 
\begin{equation}\label{eq5}
P(|\psi\rangle\rightarrow|\phi\rangle) = \min\limits_{l\in[1, n]}\left(\sum\limits_{i=l}^{n}\alpha_i\middle/\sum\limits_{i=l}^{n}\beta_i\right).
\end{equation}
For the conversion of a two-qubit pure partially entangled state with Schmidt coefficients $\sqrt{\phi_0}$ and $\sqrt{\phi_1}$, $\phi_0 > \phi_1 > 0$, to a maximally entangled Bell state, the above formula yields an SCP of 2$\phi_1$.

\subsection{Locally converting generalized GHZ to GHZ state}
We find here an LOCC-based  strategy to convert an $m$-qubit partially entangled Greenberger-Horne-Zeilinger (GHZ) state \cite{eiTa-GHZ} to the $m$-qubit GHZ state with maximal probability. The initial state for the measurement strategy is the partially entangled GHZ state (also called the generalized GHZ state), $|\psi\rangle_{A_1A_2\ldots A_m}$ = $\cos\theta|00\ldots 0\rangle+\sin\theta|11\ldots1\rangle$, $|ii\ldots i\rangle$ $\equiv$ $|i\rangle^{\otimes m}$, $\forall i = 0, 1$. Here we take $0<\theta<\frac{\pi}{4}$ and $\cos\theta$ = $\sqrt{\phi_0}>\sin\theta$ = $\sqrt{\phi_1}$. \(|0\rangle\) and \(|1\rangle\) are elements of the computational basis, being eigenstates of the Pauli-\(z\) operator. We now apply an LOCC-based measurement strategy to convert the above state to the GHZ state, $|\psi_+\rangle = (|00\ldots0\rangle + |11\ldots1\rangle)/\sqrt{2}$. The strategy involves a measurement on just any one of the $m$ qubits. The corresponding measurement operators are given by
\begin{equation}\label{eq6}
M_1 = \begin{pmatrix} \sqrt{\frac{\phi_1}{\phi_0}} & 0 \\ 0 & 1 \end{pmatrix},
\\~~~
M_2 = \begin{pmatrix} \sqrt{1-\frac{\phi_1}{\phi_0}} & 0 \\ 0 & 0 \end{pmatrix},
\end{equation}
where $\sum_{i=1}^2M_i^\dagger M_i = I$, with $I$ being the identity operator acting on the qubit Hilbert space. The probability of conversion is seen to be $2\phi_1$, and is the same as that of the conversion of the same states in any bipartition, so that the probability is optimal. 

In this paper, for \(m=3\),  we will refer to the corresponding states as generalized GHZ and GHZ states. For larger \(m\), we will refer to the GHZ state as the ``cat'' state \cite{marjar, eiTa-GHZ}.

\section{Main Results}\label{sec3}
We first demonstrate our measurement strategy using a single-layered honeycomb lattice, and then in a later portion, we extend our strategy to a double-layered honeycomb lattice by introducing entanglement percolation under restricted local operations and classical communication, i.e., rLOCC.

\subsection{Our measurement strategy on a single-layered honeycomb lattice}
Our task is to establish maximally entangled states between two distant nodes of an asymptotically large lattice. As shown in Ref.~\cite{Acin07}, the average of the SCPs over all four possible outcomes at one node that may result due to entanglement swapping between two identical copies of a two-qubit state, is the same as that of the original states. Further, they used this result to convert two layers of partially entangled two-qubit pure states arranged on honeycomb lattices to a single layer of the same on a triangular lattice, via entanglement swapping measurements at the nodes of the bi-layered honeycomb lattice. This is done because the percolation threshold of the honeycomb lattice is higher than that of the triangular one. If the partially entangled two-qubit pure states that acted as initial states of the bi-layered honeycomb lattice is such that the SCP is less than the amount needed to do entanglement percolation on a honeycomb lattice, the ACL entanglement-swapping-based quantum measurement strategy enables one to do entanglement percolation via ``moving'' to the triangular lattice, provided the said SCP is higher than the critical value needed for entanglement percolation on the latter lattice. See also Ref.~\cite{Perseguers10}.

Starting with a single layer of partially entangled pure states arranged on a honeycomb lattice, we propose another quantum measurement strategy which can attain entanglement percolation. Since, it is a single layer of the lattice that we use, the amount of entanglement used here is lower than that in the ACL strategy which used two layers. However, while ACL's strategy required two-qubit measurements, we use three-qubit ones. Later, when we use a double-layered honeycomb lattice, the single-layered strategy gets modified accordingly.

\begin{figure}[h!]
\includegraphics[scale=0.73]{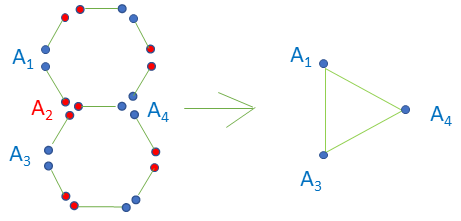}
\caption{Monolayer hexagonal lattice of bipartite states to monolayer triangular lattice of tripartite states. The single-layer hexagonal lattice is formed by bipartite (possibly non-maximally) entangled states on each edge. Each node contains three qubits. Being a bipartite lattice, we can color the nodes of the hexagonal lattice with two colors, say, red and blue, such that each nearest neighbor of a red node is blue and vice versa.  Measurement in the GHZ basis carried out on the three qubits at the red nodes. For example, the GHZ-basis measurement is performed at the node \(A_2\), which has three qubits, each of which are connected to a qubit at a neighboring blue node via a bipartite entangled state. These blue nodes are denoted in the figure as \(A_1\), \(A_3\), and \(A_4\). The relevant three qubits of these blue nodes transform into a generalized GHZ state, due to the GHZ-basis measurement at the red node, \(A_2\). There are two more qubits at each of these blue nodes, which are in turn connected to neighboring red nodes on the other side with respect to \(A_2\). Making such GHZ-basis measurements at all the red nodes of the hexagonal lattice leads to a monolayer triangular lattice of generalized GHZ states, a unit cell of which is depicted on the right-hand-side of the figure.}\label{fig3}
\end{figure}

\begin{figure}[t!]
\includegraphics[scale=0.35]{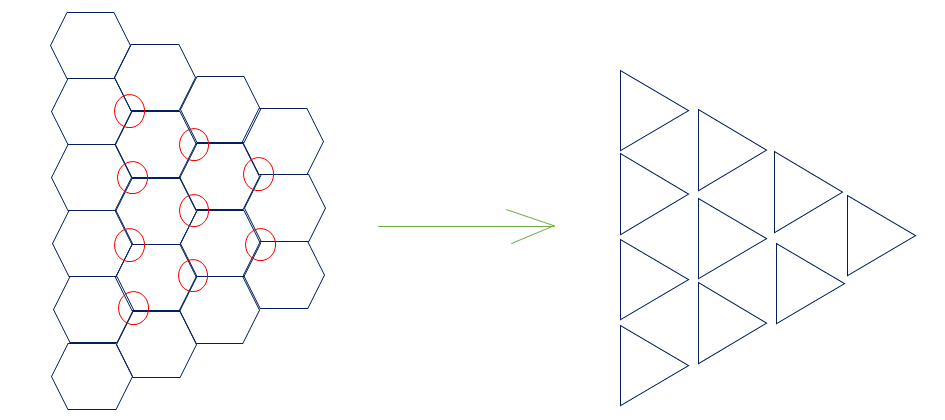}
\caption{Transforming a honeycomb lattice to a triangular one by measuring on the Greenberger-Horne-Zeilinger basis on every other node. The nodes at which the measurements are carried out are marked with a red circle on the hexagonal lattice. These nodes do not appear any more on the triangular lattice. Further details appear in the text and in the caption of Fig.~\ref{fig3}.}\label{fig4}
\end{figure}

The honeycomb lattice has three qubits at each node, and the three edges emerging from each node connect with three other qubits of three neighbouring nodes, with each connection (edge) being a single copy of the partially entangled state, $|\phi \rangle$ = $\sqrt{\phi_0}|00\rangle + \sqrt{\phi_1}|11\rangle$, $\phi_0>\phi_1 > 0$, and $\phi_0 + \phi_1 = 1$. See Fig.~\ref{fig3}. We now assume that measurements are carried out, at a certain specified set of nodes of the lattice, in the three-qubit GHZ basis which is composed of the following eight states:
\begin{equation}\label{eq7}
\begin{array}{c} 
(|000\rangle\pm|111\rangle)/\sqrt{2},~(|001\rangle\pm|110\rangle)/\sqrt{2},\\[1 ex]
(|010\rangle\pm|101\rangle)/\sqrt{2},~(|011\rangle\pm|100\rangle)/\sqrt{2}.
\end{array}
\end{equation}
The honeycomb lattice is a ``bipartite'' lattice, which means that its nodes can be colored by using two colors, say red and blue, such that all  nearest neighbors of any red node are blue, and vice versa. The GHZ-basis measurements are carried out only at the nodes of a specific color, say red. The measurements are carried out on the three qubits at the nodes which are colored red in Fig.~\ref{fig3}, and which are circled in red in Fig.~\ref{fig4}. After this measurement, we have successfully converted our single-layered honeycomb lattice made up of partially entangled pure two-qubit states to a triangular lattice spanned by three-qubit generalized GHZ states. It is to be noted that the triangular lattice is such that every fundamental triangle that is filled with a generalized GHZ state is surrounded by three empty triangles that are neighbors on its sides. And vis-{\`a}-vis, every  empty triangle is surrounded on its sides by filled triangles. 
The three-qubit generalized GHZ state between the three relevant qubits of the three nodes (which form a triangle in Fig.~\ref{fig3}) neighboring the node at which the GHZ-basis measurement is carried out, can be obtained by computing the following expression:
\begin{equation}\label{eq8}
(I\otimes|\mathcal{A}_i\rangle\langle\mathcal{A}_i|\otimes I\otimes I)|\lambda\rangle/\sqrt{p_i}.
\end{equation}
The notations in the above state can be described as follows. Consider four parties $A_1, A_2, A_3, A_4$ which are four neighboring nodes of a honeycomb lattice. Suppose that $A_2$ is sharing three pure two-qubit partially entangled states $\sqrt{\phi_0}|00\rangle+\sqrt{\phi_1}|11\rangle$, $\phi_0>\phi_1$, $\phi_0+\phi_1 = 1$, with each of the other three parties. So, $A_2$ has three qubits, representing a node, on which the measurement in GHZ basis is carried out. The operators $|\mathcal{A}_i\rangle\langle\mathcal{A}_i|$ are the projectors onto the elements of the three-qubit GHZ basis, which act on the three qubits present in a node (being in possession of \(A_2\)). On the other hand, identity operators act on the single qubits of \(A_1\), \(A_3\), and \(A_4\).  $p_i$ is the probability that the projector \(|\mathcal{A}_i \rangle \langle \mathcal{A}_i |\) clicks. \(|\lambda\rangle\) denotes the total state of the six qubits in possession of the four observers at the four nodes. Exploiting the three-qubit measurement, we obtain the following three-qubit generalized GHZ states, which generate the triangles spanning the new lattice:
\begin{equation}\label{eq9}
\begin{array}{c}
\frac{\phi_0\sqrt{\phi_0}|000\rangle\pm\phi_1\sqrt{\phi_1}|111\rangle}{\sqrt{\phi_0^3 + \phi_1^3}},~~
\frac{\phi_0\sqrt{\phi_1}|001\rangle\pm\phi_1\sqrt{\phi_0}|110\rangle}{\sqrt{\phi_0^2\phi_1 + \phi_1^2\phi_0}},\\[2 ex]
\frac{\phi_0\sqrt{\phi_1}|010\rangle\pm\phi_1\sqrt{\phi_0}|101\rangle}{\sqrt{\phi_0^2\phi_1 + \phi_1^2\phi_0}},~~
\frac{\phi_1\sqrt{\phi_0}|011\rangle\pm\phi_0\sqrt{\phi_1}|100\rangle}{\sqrt{\phi_0^2\phi_1 + \phi_1^2\phi_0}}.
\end{array}
\end{equation}
These generalized GHZ states are created due to a GHZ-basis measurement, and they appear, respectively, with probabilities \(p_i\), given by 
\begin{eqnarray}
&& p_1 = p_2 = \frac{\phi_0^3 + \phi_1^3}{2},\nonumber \\
&& p_j = \frac{\phi_0^2 \phi_1 + \phi_1^2 \phi_0}{2},~\forall j = 3, \ldots, 8. 
\end{eqnarray}

\begin{figure}[b!]
\includegraphics[scale=0.335]{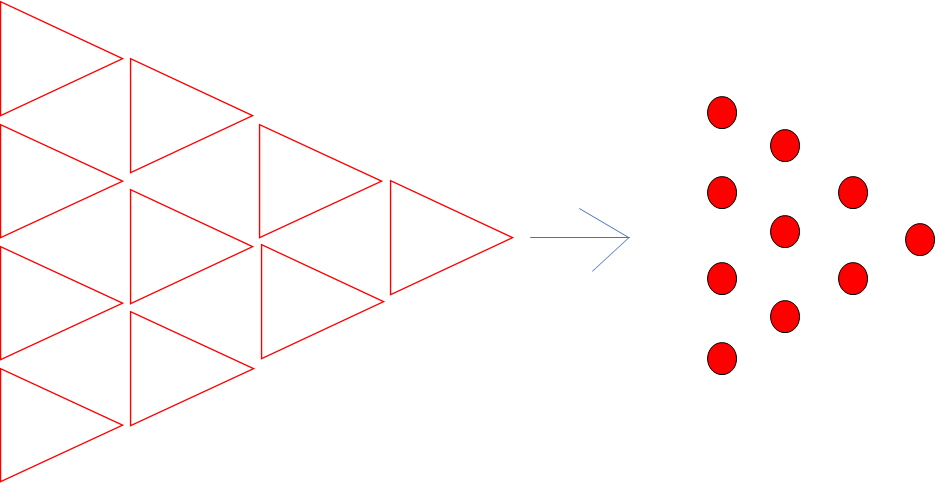}
\caption{Mapping bond percolation problem to site percolation one. The triangular lattice of generalized GHZ states that we obtained via the GHZ-basis measurements on every other node of the hexagonal lattice is depicted on the left-hand-side of the figure. The intent is to create a cat state (see text) between qubits of an arbitrarily large number of nodes. On the left-hand-side, this is a bond percolation problem, while we can look at it as a site percolation problem by replacing every generalized GHZ states on the left panel by a dot on the right panel. Each of these dots has a three-qubit GHZ state with probability \(p_0\). The same intent as in the left panel is attained by a site percolation on the triangular  lattice on the right panel. The change is only at the level of calculations, and does not require a physical transformation.}\label{fig5}
\end{figure}

The average SCP is calculated by averaging the SCPs over all the eight possible outcomes (all outcomes are given above) and is given as Avg.~SCP = $p_0$ = $2\phi_1^2(\phi_1+3\phi_0)$. Now we need to figure out the percolation threshold for our triangular lattice. It could be difficult to calculate the threshold for bond percolation in a triangular lattice spanned by GHZ states. However, we can map our problem to a site percolation problem as shown in Fig.~\ref{fig5}.

\begin{figure}[h!]
\includegraphics[scale=0.6]{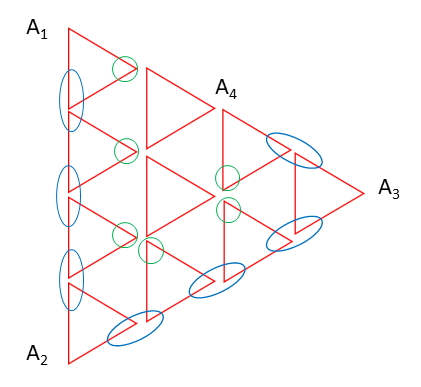}
\caption{A specific scenario of percolation on triangular lattice of Greenberger-Horne-Zeilinger states. Suppose that we wish to create a cat state (see text) between the nodes marked as \(A_1\), ... \(A_4\). We are beginning with an initial situation where each triangle in the figure represents a GHZ state. For every such triangle, the GHZ state is present with probability \(p_0\). The cat state that we wished for, can be obtained by Bell measurements and single-qubit measurements along the boundary of the region formed by the nodes \(A_1\), ... \(A_4\). The Bell measurements are performed at the nodes marked by blue ellipses, on the two qubits inside those ellipses. The single-qubit measurements are marked by green circles.}\label{fig6}
\end{figure}

An essential point to note here is that now each site in the mapped triangular lattice (the red dots on the right-hand-side lattice in Fig. \ref{fig5}) denotes the presence of a GHZ state with probability \(p_0\)  (and its absence with probability \(1-p_0\)). Percolation of GHZ states in the original lattice is mapped to percolation of sites in the mapped triangular lattice of sites in a one-to-one correspondence. It is to be noted that the mapping here is just a mental picture that aids in the mathematics of the problem, and does not represent a physical manoeuvre. 

The site percolation threshold for a triangular lattice is \(1/2\) \cite{ekTa-gaan-likho-amar-janye}. If the average SCP, \(p_0\), of the original triangular lattice with generalized GHZ states (left panel in Fig. \ref{fig5}) is larger than the percolation threshold \(p_{c_{\Delta}} \equiv 1/2\) of the mapped triangular lattice of sites (right panel in Fig. \ref{fig5}), arbitrarily large cat states, \(\frac{1}{\sqrt{2}}(|00 \ldots 0 \rangle + |11 \ldots 1\rangle)\), will be formed in the original triangular lattice. This will be effected in the following way. There is a probability \(p_0\) for the the generalized GHZ states in the original triangular lattice to be transformed locally, i.e., by local (with respect to the sites) quantum operations and classical communication (between the sites), to a GHZ state. In cases when the transformation is successful, we do two-qubit Bell-basis measurements along with single-qubit measurements at the sites of the original triangular lattice along the boundary of the region bounded by the sites forming the cat state. See Fig.~\ref{fig6} for an example. The condition, on the parameters of the bipartite non-maximally entangled states of the hexagonal lattice, for  successfully creating an arbitrarily large GHZ state is given by 
\begin{equation}\label{se-mandire-debata-nai}
2\phi_1^2\left(\phi_1+3\phi_0\right) > \frac{1}{2},
\end{equation}
which, solving the cubic, provides the range, 
\begin{equation}\label{debatar-dhan-ke-jai-phiraye-laye-eibyala-shon}
\phi_0 > \phi_1 \in \left(\frac{1}{2}-\sin\frac{\pi}{18}, \frac{1}{2}\right),
\end{equation}
that is,
\begin{equation}\label{debatar-dhan-ke-jai-phiraye-laye-eibyala-shon}
\phi_0 > \phi_1 \in \left(0.32635, \frac{1}{2}\right),
\end{equation}
with the left end of the last interval being correct to five significant figures. This threshold is exactly the same as for entanglement percolation using CEP on the hexagonal lattice \cite{Acin07}. Each site of the original triangular lattice contains three qubits. However, the sites that will form the cat state will of course have  one ``active'' qubit (i.e., the qubit used in the construction of the cat state), and the remaining two qubits will remain ``passive''. By measuring in the \(\sigma_x\)-basis on the active qubits at all but two of these sites, we can create a maximally entangled bipartite state between two sites that have an arbitrarily large distance between them on the lattice. It is evident that our measurement strategy, when applied on a single-layered honeycomb lattice, is not better than CEP. However, our measurement strategy helps to reduce the number of measurements. We now proceed to talk about that issue.

{\it Reducing the number of measurements}.-- Applying our measurement strategy to a single-layered honeycomb lattice for the percolation task, reduces the number of required measurements, providing potentially important implications for noise resistance in an actual realization of our technique.

Comparing our result with that in Ref.~\cite{Acin07}, we see that there the authors converted the bilayer honeycomb lattice of partially entangled two-qubit states to a triangular lattice, and they succeeded in attaining bipartite entanglement percolation. All measurements performed were two-qubit ones. Our measurement strategy applied on a single layer of their honeycomb lattice uses three-, two-, and single-qubit measurements, to attain multipartite (and hence also bipartite) entanglement percolation. 

We compare our result of using QEP on the monolayer hexagonal lattice with that of using CEP on the same lattice~\cite{Acin07}. The thresholds obtained are exactly the same. However, the number of measurements  are different. While CEP uses lower-qubit measurements, we use less measurements. Precisely, for an \(l \times l\) square box encompassing a part of the hexagonal lattice, to create a cat state between nodes on the boundary of the square, CEP requires \(6l^2 + O(l)\) single-qubit and \(O(l)\) two-qubit measurements, while the QEP proposed here requires \(2l^2 + O(l)\) single-qubit, \(O(l)\) two-qubit, and \(2l^2 + O(l)\) three-qubit measurements. It is to be noted that the length and breadth of the square box are counted such that the hexagonal lattice in Fig.~\ref{fig4} is of breadth two. It is probably important to mention the following here. If the threshold required to connect nodes agrees with the one given through the percolation threshold then it is expected that a connection between the distant ends can be established for some nodes. However, it cannot be decided in advance which nodes at the distant ends would be connected in the end. To see the difference more clearly, we highlight the comparison in the appended table.
\begin{table} [h!]
\centering
\begin{tabular}{|c|c|c|}
\hline
\textbf{Type of Measurement} & \textbf{QEP} & \textbf{CEP} \\
\hline
\textbf{Single-qubit measurements} & $2l^2 + O(l)$ & $6l^2 + O(l)$  \\
\textbf{Two-qubit measurements} & $O(l)$ & $O(l)$ \\
\textbf{Three-qubit measurements} & $2l^2 + O(l)$ & -\\
\hline
\hline
\textbf{Total Number of Measurements} & $4l^2 + O(l) $ & $6l^2 + O(l)$\\
\hline
\end{tabular}
\caption{Comparison of the number of measurements in the strategy described in this paper with the CEP in Ref.~\cite{Acin07}. The number of measurements shown here are for an $l \times l$  square box encompassing a part of a hexagonal lattice. For a large lattice, the number of measurements that our strategy uses would be considerably lower than in CEP, and hence it will potentially call for less noise effects, assuming a physical system that has comparable noise levels in single-, and two-, and three-qubit measurements.}\label{tab1}
\end{table}

Finally, in Ref.~\cite{Perseguers10}, the authors have used a different measurement strategy and have attained  multipartite entanglement percolation. For the success of their strategy, the authors needed to do between four and five measurements per unit cell of their honeycomb lattice, whereas we need to do three measurements per unit cell of our honeycomb lattice. Due to the reduced number of measurements (per unit cell),  it is plausible that noise effects on a realization of our strategy will be lower than the same on the one in Ref.~\cite{Perseguers10}. 

\subsection{Our measurement strategy on a double-layered honeycomb lattice}
We have already seen that using a single-layered honeycomb lattice, our measurement strategy does not lend any advantage over CEP. So, it may seem that our measurement strategy is weaker than the one introduced in~\cite{Acin07}. But to explore the non-trivial advantage of our measurement strategy we need to dig a bit deeper. Though till now we have just used a single layer of a honeycomb lattice, we can show that our measurement is of particular importance when there are more than one layers. The same measurement strategy, as introduced in the previous section, is extended for the double-layered honeycomb lattice here. As in case of \cite{Acin07}, the two layers of our double-layered honeycomb lattice are built of non-maximally entangled pure quantum states. But in this network, only restricted local operations and classical communication (rLOCC) is allowed. We now provide a formal definition of rLOCC.

\begin{definition}
The LOCC class that we had considered until now meant that an observer at a given node can perform all quantum operations at that node, and can also communicate classically with the observers at any other node. The restricted LOCC class disallows any operation in that class that involves a joint  operation on ``particles'' of  both the layers. The parties must perform LOCC between the nodes, but by considering one layer after another.
\end{definition}

The motivation of considering rLOCC comes from the concept of adaptive LOCC. Researchers often use adaptive LOCC in the context of state discrimination or channel discrimination in many-copy scenario. Adaptive LOCC is interesting because it is potentially easier to  implement in a practical scenario, in comparison to an element of the full LOCC class. This is the reason why we consider rLOCC.

We next consider $|\phi \rangle ^{\otimes 2}$, where again $|\phi \rangle$ = $\sqrt{\phi_0}|00\rangle + \sqrt{\phi_1}|11\rangle$, $\phi_0>\phi_1 > 0$, and $\phi_0 + \phi_1 = 1$. We also consider that each of the $|\phi \rangle$s  belong to the two different layers. So, the optimal singlet conversion probability (SCP) under rLOCC will be $p + (1 - p)p$, where $p = 2\phi_1.$ Given the first copy, the parties are able to succeed with probability $p = 2\phi_1$. But they fail with probability $(1 - p)$. So, with probability $(1-p)$, they use the second copy and then they can succeed with probability $p = 2\phi_1$ again. In this way, the overall optimal SCP under rLOCC is given by $p + (1 - p)p$. For CEP to succeed we need,
\begin{equation}
p + (1-p)p > \left( 1- 2\sin\frac{\pi}{18}\right)    
\end{equation}
which implies that,
\begin{equation}
\begin{split}
\phi_1 > \frac{1}{2}\left(1 - \sqrt{\sin\frac{\pi}{18}}\right) \approx 0.2916 \\
\implies \phi_1 > 0.2916
\end{split}
\end{equation}
Clearly, if we take $\phi=0.28$, then CEP fails. Nevertheless for this choice of $\phi_1=0.28$, the average SCP using our measurement strategy as introduced in the last section is 
\begin{equation}
p'= 2\phi_1^2(\phi_1 + 3\phi_0),
\end{equation}
that is,
\begin{equation}
\begin{split}
p'= 0.3825.
\end{split}
\end{equation}
Recall that $p'$ is achievable when there is only one layer. But here we are considering two layers under rLOCC. In this case, using the first layer, probability $p'$ is achievable. Then, with probability $(1 - p')$, the parties fail and they use the second layer. Again, they get success with probability $p'$. So, the overall optimal SCP under rLOCC using our measurement strategy is $p'+(1-p')p'$ $= 0.6186 > 0.5$, with $p'=0.3825$ and where the site percolation threshold for the triangular lattice is $0.5$. So, clearly QEP using our measurement strategy succeeds where CEP alone had failed.

In the above, we discussed an example where we started with $\phi_1$ = $0.28$, i.e., the corresponding non-maximally entangled state is given by- $\ket{\phi}$ = $\sqrt{0.72}\ket{00}+\sqrt{0.28}\ket{11}$. For the layers with this non-maximally entangled state, we saw that when we use rLOCC, CEP fails but QEP works. More precisely, under rLOCC, the bound of (15) is strict for CEP, but if one uses QEP, then, it is possible to go beyond this bound.  

The above is interesting due to the following reasons:
\begin{itemize}
\item We have introduced the concept of entanglement percolation under rLOCC.

\item rLOCC is motivated from the real-life scenarios. Implementation of rLOCC must be easier compared to full LOCC.

\item Another important point is that if we look into the protocol of \cite{Acin07}, then under rLOCC, their protocol fails. That is because they use measurements in the Bell basis on both the layers jointly. On the contrary our measurement strategy succeeds under rLOCC.

\item When we do not get an advantage of QEP over CEP using our strategy, apparently, it seems that our strategy is weak. However, we find an advantage when there are more than one layers and under rLOCC.

\item Lastly, in the context of entanglement percolation, we find a nontrivial feature in a many-layer scenario which is not occurring in the single layer scenario.
\end{itemize}

\section{Conclusion}\label{sec4}
Entanglement percolation is an interesting technique to distribute entangled states between two or more nodes of a lattice that can be arbitrarily distant. It is important to note that it cannot be decided in advance as to which nodes at the distant ends would be connected in the end, but the important thing is that such a connection between the distant ends can be established for some nodes. We have used both single- and double-layered honeycomb lattices, made of non-maximally entangled pure bipartite quantum states. We have provided a quantum measurement strategy involving three-, two-, and single-qubit measurements, to obtain Greenberger-Horne-Zeilinger (cat) states shared between an arbitrarily large number of lattice nodes. The cat states can then be reduced to a two-qubit Bell state shared between faraway nodes. 

A feature of our strategy is that after our entanglement swapping measurement on the initial hexagonal lattice, every other node with all their qubits is totally removed from the protocol and does not play any further role in the strategy, which is significantly different from the measurement strategy used in Ref.~\cite{Perseguers10}. The resources used in an entanglement percolation strategy are the entangled states of the original lattice and the measurements performed in between. The number of measurements performed is potentially an important parameter for estimating the noise effects on a realization of the strategy. We have compared our entanglement percolation strategy with existing ones in the literature with respect to both these resources and our result shows that an experimental realization based on our measurement strategy will potentially have lower noise effects than in those based on the strategies in Refs.~\cite{Acin07,Perseguers10}. This is when we use a single-layered honeycomb lattice.

Moreover, our measurement strategy is effective for entanglement percolation under rLOCC - an experiment-friendly restriction to the LOCC class - in a double-layered honeycomb lattice. In particular, we have reported advantage of quantum entanglement percolation over its classical counterpart under the restricted class of operations. Under the same restriction, the strategy of \cite{Acin07} does not succeed in realizing entanglement percolation.

\end{document}